\documentclass[cameraready]{Interspeech}

\title{An Efficient vLLM-Based Inference Pipeline for Unified Audio Understanding and Generation}

\author[affiliation={1,2}]{Haoran}{Wang}
\author[affiliation={1}]{Jinchuan}{Tian}
\author[affiliation={1}]{Siddhant}{Arora}
\author[affiliation={1}, correspondingauthor]{Shinji}{Watanabe}

\address{
    $^1$ Carnegie Mellon University \\
    $^2$ Shanghai Jiao Tong University
}

\email{haoranw5@andrew.cmu.edu}


\usepackage{comment}
\usepackage{graphicx}
\usepackage{hyperref}
\usepackage{booktabs}
\usepackage{makecell}
\usepackage{enumitem}
\usepackage{multirow}

\begin{document}

\maketitle

\begin{abstract}
    While Large Multimodal Models excel in comprehension, high-throughput inference engines lack native support for multimodal generation. This is severe in Speech Language Models, where generating multi-layered audio tokens via decoupled AR+NAR or synchronous Multi-Token Prediction (MTP) with delay-pattern interleaving conflicts with standard single-stream loops. We present a vLLM-based inference pipeline for unified speech understanding and generation. We extend autoregressive decoding to natively execute delay-pattern de-interleaving and coordinated multi-stream sampling, integrating an on-GPU acoustic decoder for end-to-end waveform synthesis. Crucially, we overcome the shared intuition that Classifier-Free Guidance (CFG) halves throughput. By co-scheduling paired conditional and unconditional requests within a continuous batch, our CFG implementation sustains 80\% of non-CFG throughput, absorbing dual-request and logit merging overheads. We open-source our framework\footnote{\url{https://github.com/whr-a/vllm/tree/opuslm}}.
\end{abstract}

\section{Introduction}

The rapid evolution of Large Language Models (LLMs) has transitioned artificial intelligence from text-only interfaces to rich, multimodal interactions. Recently, Speech Language Models (SpeechLMs)~\cite{tian2026bagpiper, tian2025opuslm, xu2025qwen3, ding2025kimi} have emerged as a pivotal frontier for unified audio understanding and generation. By modeling continuous audio signals as discrete acoustic tokens~\cite{shi2024espnet, ye2025codec, wang2025bscodec}, these models bypass traditional cascaded Automatic Speech Recognition (ASR)~\cite{radford2023robust} and Text-to-Speech (TTS)~\cite{kim2021conditional, shen2018natural} pipelines to achieve strong and expressive acoustic capabilities. Building upon this end-to-end foundation, advanced interactive features like direct speech-to-speech dialogue~\cite{arora2026optimizing, defossez2024moshi} naturally emerge as downstream applications.

However, the deployment of SpeechLMs exposes a structural gap in current high-throughput inference engines. While state-of-the-art serving frameworks~\cite{kwon2023efficient, zheng2024sglang} excel at memory efficiency and dynamic request scheduling (e.g., continuous batching~\cite{yu2022orca} and automatic prefix caching~\cite{zheng2024sglang}), their underlying mechanisms were largely optimized for unimodal, text-only generation. Consequently, extending this text-centric architecture to handle the continuous and distinct nature of speech outputs introduces significant bottlenecks.

Generating general audio diverges from standard text autoregression in two critical aspects:
\begin{enumerate}
    
    \item \textbf{Multi-Layered Token Generation and Decoding.} Unlike text, continuous speech is typically compressed by neural audio codecs~\cite{shi2024espnet, wang2025bscodec, mousavi2025discrete, guo2025recent}. Modern SpeechLMs favor Residual Vector Quantization (RVQ), encoding each temporal frame into multi-layered tokens to support high-fidelity audio at a constant frame rate. To model this multi-codebook grid, current paradigms primarily adopt synchronous Multi-Token Prediction (MTP)~\cite{xu2025qwen3} and delay-pattern interleaving~\cite{tian2026bagpiper, copet2023simple}. While delay-pattern elegantly flattens tokens for training, its inference demands multi-stream sampling and complex de-interleaving. Furthermore, synthesizing the final waveform requires an auxiliary acoustic decoder (vocoder). Coupling MTP with this additional decoding stage inherently conflicts with the standard one-token-per-step serving loop.
    
    \item \textbf{The Classifier-Free Guidance (CFG) Bottleneck.} Classifier-Free Guidance (CFG)~\cite{ho2022classifier} is essential for high-fidelity audio generation, as evidenced by its widespread adoption in various audio synthesis frameworks~\cite{copet2023simple, tian2025ualm, liu2023audioldm}. Mechanistically, implementing CFG requires two parallel forward passes per step: one conditioned on the input prompt and one unconditional. This dual-pass requirement inherently doubles computational and memory overheads. Since standard schedulers assume strict request independence, they lack native mechanisms to coordinate these paired passes. Consequently, naive implementations are forced to process them as separate requests. This disjoint scheduling duplicates KV cache allocations and precludes joint batching. Furthermore, the necessary logit merging must operate across disjoint request contexts with explicit synchronization. The compounding of these dual-request overheads and synchronization costs can severely degrade overall system throughput.
\end{enumerate}

To bridge this gap, we present a novel inference implementation engineered for the unified understanding and generation of audio. Built upon vLLM~\cite{kwon2023efficient} which is a high-performance continuous batching backend, our framework extends the autoregressive decoding loop to natively support synchronous multi-stream sampling with delay-pattern de-interleaving for multi-codebook audio generation. Furthermore, we fuse a lightweight acoustic decoder directly into the engine's output pipeline, enabling single-process audio synthesis without an external vocoder service

To address the CFG overhead, we introduce a Paired Request Co-Scheduling mechanism. By co-scheduling paired conditional and unconditional requests within the same continuous batch, our approach ensures them to share a single forward pass through the transformer backbone. This unified execution elegantly absorbs the dual-request and logits merging overheads, directly overcoming the shared intuition that CFG inevitably halves system throughput.

The main contributions are summarized as follows.
\begin{enumerate}
    \item We identify the structural limitations of current text-centric inference engines in handling multimodal auto-regressive generation and propose a unified architecture that natively supports synchronous multi-stream sampling and delay-pattern de-interleaving for multi-codebook audio tokens.

    \item We design an end-to-end fused acoustic pipeline that integrates a lightweight audio decoder directly into the engine's output pipeline, eliminating the need for an external vocoder service and enabling single-process waveform synthesis on GPU.

    \item We introduce a Paired Request Co-Scheduling strategy for CFG. By systematically pairing and co-scheduling conditional and unconditional requests within the same continuous batch, our system efficiently absorbs dual-branch overheads and logit synchronization costs, sustaining up to 80\% of non-CFG throughput and effectively overcoming the shared intuition of halved performance, while seamlessly supporting the concurrent prefilling and decoding of text and CFG-enabled audio requests in a unified pipeline.
\end{enumerate}

\section{Method}
We present a serving framework that enables high-throughput inference for unified audio understanding and generation within existing token-level continuous batching engines. We first describe the class of models our framework targets (\S\ref{sec:method:model}), then detail our unified serving pipeline (\S\ref{sec:method:pipeline}), and finally describe efficient Classifier-Free Guidance (\S\ref{sec:method:cfg}).

\begin{figure}[t]
    \centering
    \includegraphics[width=\columnwidth,trim={18pt 0 30pt 18pt},clip]{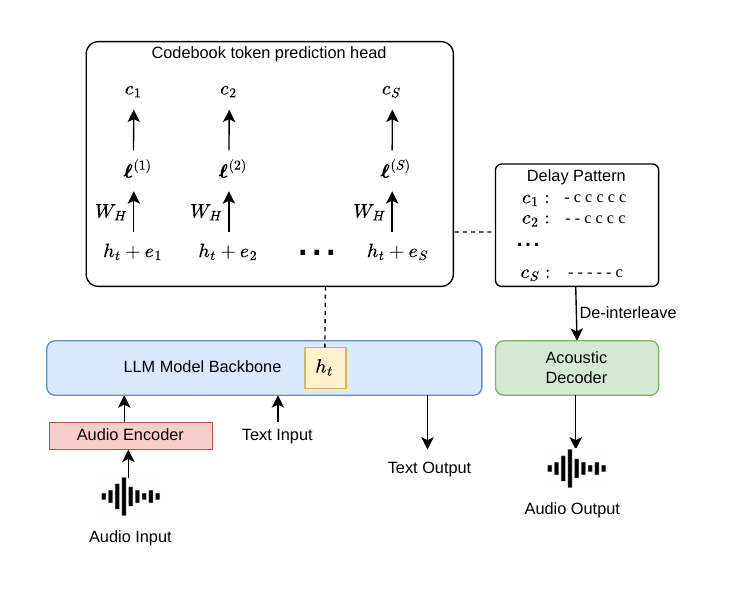}
    \vspace{-15pt}
    \caption{Speech language model architecture. The unified autoregressive transformer processes encoded audio and text. During audio generation, hidden states $h_t$ and stream embeddings $e_i$ (where $1 \le i \le S$) are projected via $S$ codebook heads into parallel delayed streams, which are subsequently de-interleaved for acoustic decoding.}
    \vspace{-15pt}
    \label{fig:model}
\end{figure}

\subsection{Speech Language Models}
\label{sec:method:model}

Our system broadly targets the widely adopted \emph{delay-pattern}~\cite{copet2023simple} SpeechLM architecture. To validate the generality of our pipeline, we ground our implementation and evaluation in three representative models: Bagpiper~\cite{tian2026bagpiper}, OpusLM~\cite{tian2025opuslm}, and OpusLM dialogue~\cite{arora2026optimizing}. As illustrated in Figure~\ref{fig:model}, the core generation mechanisms of these models share a unified architectural foundation.

\textbf{Multi-token Prediction.} The pipeline begins with an audio encoder feeding continuous speech to an autoregressive LLM. The LLM operates over a unified vocabulary, partitioned into text and $S$ audio sub-vocabulary. For audio output, at each time step $t$, the LLM backbone produces a hidden state $h_t$. As illustrated in Figure~\ref{fig:model}, $S$ parallel codebook heads then project this state to generate logits $\boldsymbol{\ell}^{(i)}$ and their corresponding sampled tokens $c_i$ (where $1 \le i \le S$). Specifically, each head adds a learnable stream embedding $e_i$ to $h_t$ before applying a shared language modeling head $W_{H}$, such that $\boldsymbol{\ell}^{(i)} = W_{H}(h_t + e_i)$.

\textbf{Delay Interleave pattern.} Under the delay pattern~\cite{copet2023simple}, the $i$-th stream is offset by $i$ steps, meaning the $S$ tokens $\{c_1, \dots, c_S\}$ emitted at step $t$ correspond to different temporal positions. To maintain this structure during autoregressive feedback, the input representation at each step aggregates the embeddings of these previously generated tokens. Post-generation, the resulting $S \times T$ codebook matrix (where $T$ represents the total number of acoustic frames) is de-interleaved and passed to an acoustic decoder for waveform synthesis. Crucially, this unified vocabulary inherently enables mixed-modality responses (e.g., text seamlessly transitioning to audio).

\subsection{Standard Continuous Batching Engines}
\label{sec:method:serving_engine}

Modern text-based LLM serving frameworks, such as vLLM~\cite{kwon2023efficient}, maximize hardware utilization through iteration-level scheduling and continuous batching. Rather than processing static batches where early-finishing requests idle, the scheduler dynamically injects new prompts and retires completed generations at the granularity of a single decoding step. This highly dynamic scheduling is supported by memory management mechanisms like PagedAttention~\cite{kwon2023efficient}, which partitions the Key-Value (KV) cache into fixed-size, non-contiguous blocks to allocate GPU memory on demand and eliminate fragmentation. Structurally, these engines treat the underlying model as a black box that executes a uniform forward pass across all active requests. Consequently, the entire system infrastructure is deeply coupled to the assumption of homogeneous, single-stream text generation, presenting a strict architectural constraint that our framework must navigate to support multi-token prediction and CFG inference.

\subsection{Unified Multi-Token Prediction Serving Pipeline}
\label{sec:method:pipeline}

To navigate this single-stream constraint and support $S$ parallel audio tokens without requiring pervasive modifications to the engine's iteration-level scheduler or PagedAttention~\cite{kwon2023efficient} infrastructure, we introduce a \emph{primary-auxiliary decomposition}, as depicted in Figure~\ref{fig:overall} (Left).

\textbf{Primary-Auxiliary Decomposition.} We designate one codebook stream as the \emph{primary} token, which is fully managed by the engine's standard pipeline. The remaining $S{-}1$ \emph{auxiliary} streams are sampled internally during the model's logit computation and buffered per-request. Consequently, the model appears to the engine as a standard text LM that emits a single token per step, while transparently executing the multi-codebook generation.

\textbf{De-interleave and Decode.} Upon request completion, the buffered matrix is delay-de-interleaved and passed to a co-located on-GPU acoustic decoder, fusing waveform synthesis directly into the serving path.

\textbf{Mixed-Modality Phase Management.} To orchestrate mixed text-audio outputs within this token-level engine, we implement a per-request \emph{phase state machine} enforced via dynamic vocabulary masking. The system transitions across three main states: a \emph{text phase} restricted to text tokens, a \emph{transition phase} injecting deterministic bridging tokens to condition audio generation, and an \emph{audio phase} activating all $S$ streams. Crucially, terminating the audio phase initiates a \emph{drain sub-phase} that emits $S{-}1$ padding tokens to flush the delay buffer, guaranteeing a complete $S \times T$ codebook matrix (where $T$ represents the total number of acoustic frames) for de-interleaving. Furthermore, because continuous batching interleaves tokens from multiple requests into a single flattened sequence, tracking modality via input positions is unreliable. We bypass this by dynamically inferring the active modality directly from the model's unmasked output distribution (i.e., the argmax over the disjoint vocabulary regions), ensuring robust and engine-agnostic phase management.

\subsection{Classifier-Free Guidance via Paired Request Co-Scheduling}
\label{sec:method:cfg}

\begin{figure}[t]
    \centering
    \includegraphics[width=\columnwidth,trim={0 0 0 15pt},clip]{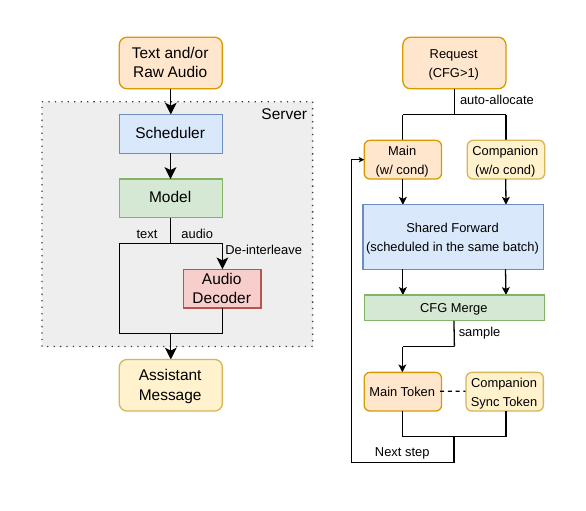}
    \vspace{-20pt}
    \caption{Unified serving framework for SpeechLMs. \textbf{(Left)} Overall pipeline: The engine manages continuous batching, primary-auxiliary token generation, and fused on-GPU acoustic decoding for mixed-modality streaming. \textbf{(Right)} Paired co-scheduling for CFG: Conditional and unconditional requests share a batched forward pass, merging logits before sampling and synchronizing the sampled token for the next step.}
    \vspace{-12pt}
\label{fig:overall}
\end{figure}


Classifier-Free Guidance (CFG)~\cite{ho2022classifier} improves audio fidelity by interpolating between conditional and unconditional distributions. Building upon the logit formulation $\boldsymbol{\ell}^{(i)}$ defined in Section~\ref{sec:method:pipeline}, we apply CFG independently to each codebook stream $i$:
\begin{equation}
    \tilde{\boldsymbol{\ell}}^{(i)} = w \cdot \boldsymbol{\ell}^{(i)}_{\text{cond}} + (1 - w) \cdot \boldsymbol{\ell}^{(i)}_{\text{uncond}}, \quad w > 1
    \label{eq:cfg}
\end{equation}
A straightforward implementation would issue two independent requests, incurring severe scheduling fragmentation and cross-request synchronization overhead, and is not feasible because it needs step-by-step sync. To resolve this, we introduce \emph{paired request co-scheduling}, which natively coordinates this dual-pass requirement within the serving engine.

As detailed in Figure~\ref{fig:overall} (Right), the end-to-end processing pipeline begins with the input auto-allocation: when the server receives a request with a CFG scale $w > 1$, it transparently generates a ``Companion'' request (with a null prompt) alongside the ``Main'' conditional request. The engine treats this pair as an atomic scheduling unit and always process them together. During the shared forward process, they are co-scheduled into the same continuous batch and share a single forward pass through the transformer backbone. The engine strictly partitions the per-step token budget equally between the pair.

After the forward pass, the raw logits $\boldsymbol{\ell}^{(i)}_{\text{cond}}$ and $\boldsymbol{\ell}^{(i)}_{\text{uncond}}$ from both contexts are merged according to Eq.~\ref{eq:cfg}. Crucially, this CFG merge operates only during the audio phase; it bypasses modality-switching tokens to prevent CFG from corrupting the phase state machine. To avoid redundant compute, projections for auxiliary streams are batched together. 

Finally, for the output and synchronization step, the engine samples a single Main Token from the merged distribution $\tilde{\boldsymbol{\ell}}^{(i)}$. To ensure both contexts maintain the exact same autoregressive history for the next step, this sampled token is synchronously appended to the Companion request, while the companion's external outputs remain hidden from the client. By systematically co-scheduling these pairs within the same continuous batch, our architecture efficiently absorbs the dual-branch overheads and logit synchronization costs. This design effectively overcomes the conventional assumption that CFG halves performance, allowing the system to sustain up to 80\% of non-CFG throughput.

\section{Experiment}

\subsection{Experimental Setup}
\label{sec:exp:setup}

We evaluate our framework on three fully open-source SpeechLM architectures that span different design points, all sharing a delay-pattern codec structure and supporting both audio understanding and generation:
\begin{itemize}[leftmargin=*,nosep]
    \item \textbf{Bagpiper}~\cite{tian2026bagpiper}: A model utilizing a Qwen3-8B~\cite{yang2025qwen3} backbone with 8 parallel codebook streams encoded via X-Codec~\cite{ye2025codec}. It is designed for open-ended audio understanding and generation, natively supporting Classifier-Free Guidance for high-fidelity audio synthesis.
    \item \textbf{OpusLM}~\cite{tian2025opuslm}: A model featuring an OLMo-2-7B~\cite{olmo20242} backbone with a 9-stream setup (1 SSL~\cite{chen2024towards} + 8 DAC~\cite{shi2024espnet, kumar2023high}), tailored for both ASR and TTS tasks.
    \item \textbf{OpusLM-Dialogue}~\cite{arora2026optimizing}: A model adopting a SmolLM2-1.7B~\cite{allal2025smollm2} backbone. It retains the same 9-stream codec configuration as OpusLM but is specifically optimized for spoken dialogue tasks.
\end{itemize}
For the baseline, we use the original PyTorch inference pipeline, which processes requests sequentially without continuous batching. All experiments are conducted on a single NVIDIA H100 80GB GPU with FlashAttention-3~\cite{shah2024flashattention} enabled. The maximum model sequence length is set to 16{,}384 tokens for all configurations. To maximize operational efficiency and fully saturate the GPU, we maintain a concurrency level of 512 simultaneous requests for standard evaluations. For CFG-enabled settings, the concurrency is proportionally adjusted to 256 requests to account for the internal companion requests, yielding an equivalent active workload. This targeted concurrency ensures that the serving engine maintains a steady, moderately sized queue to sustain peak overall throughput.

\subsection{Throughput}
\label{sec:exp:throughput}

In addition to decode token rates, we measure Model FLOPs Utilization (MFU), which quantifies the ratio of actual floating-point operations to the GPU's theoretical peak capacity, providing a standardized metric for hardware utility. As summarized in Table~\ref{tab:throughput}, our continuous-batching pipeline achieves massive performance gains over the sequential PyTorch baseline across all models. By effectively batching multi-stream generation, our system accelerates token processing speeds by roughly two orders of magnitude. This architectural advantage directly translates to hardware utilization: while the baseline severely underutilizes the H100 GPU with an MFU well below 0.1\%, our serving engine efficiently absorbs scheduling overheads to drive the MFU up to 9.95\% for Bagpiper and 9.89\% for OpusLM. Notably, the 1.7B parameter OpusLM dialogue model exhibits a lower MFU of 3.28\% despite achieving the highest absolute token throughput. This occurs because such a compact model possesses low arithmetic intensity on the H100 GPU, causing inference to hit the memory bandwidth wall before fully saturating the compute units.

\begin{table}[t]
    \centering
    \caption{Prompt and generation throughput comparison. All measurements are conducted on a single H100 80GB GPU with FlashAttention-3 and a maximum sequence length of 16{,}384.}
    \label{tab:throughput}
    \setlength{\tabcolsep}{3pt}
    \footnotesize
    \begin{tabular}{lccc}
        \toprule
        \textbf{System} & \textbf{Decode (tok/s)} & \textbf{MFU (\%)} \\
        \midrule
        Bagpiper (PyTorch) & 52.7 & 0.096 \\
        Bagpiper (vLLM) & 5694.5 & 9.95 \\
        \midrule
        OpusLM (PyTorch) & 36.5 & 0.311 \\
        OpusLM (vLLM) & 4582.9 & 9.89 \\
        \midrule
        OpusLM-Dial. (PyTorch) & 53.9 & 0.290 \\
        OpusLM-Dial. (vLLM) & 5870.5 & 3.28 \\
        \bottomrule
    \end{tabular}
    \vspace{-10pt}
\end{table}





\subsection{Numerical and Quality Check}
\label{sec:exp:numerical}

A prerequisite for any serving optimization is that it must not degrade model output quality. We verify numerical correctness using the Bagpiper model~\cite{tian2026bagpiper} on the MMAU-mini dataset~\cite{sakshi2024mmau}. Specifically, we compare the token sequences and logit distributions produced by our continuous-batching pipeline against the reference PyTorch implementation under identical inputs and random seeds.

\begin{table}[h]
    \centering
    \footnotesize
    \caption{Numerical deviation against the reference PyTorch implementation.}
    \begin{tabular}{lccc}
        \toprule
        \multirow{2}{*}{\textbf{Configuration}} & \textbf{RMSE} & \textbf{RMSE} & \textbf{First} \\
         & \textbf{(Max)} & \textbf{(Mean)} & \textbf{Mismatch} \\
        \midrule
        FP32 + SDPA~\cite{vaswani2017attention} & 0.081 & 0.008 & None (Identical) \\
        BF16 + FA3~\cite{shah2024flashattention} & 0.532 & 0.163 & 34.0 \\
        \bottomrule
    \end{tabular}
    \label{tab:numerical}
\end{table}

As shown in Table~\ref{tab:numerical}, under Float32 precision, the two implementations produce strictly identical token sequences, with a negligible mean logit RMSE of 0.008. This exact alignment confirms that our custom multi-stream sampling and phase control introduce no numerical bugs. When transitioning to BF16 with FlashAttention-3~\cite{shah2024flashattention}, a mismatch occurs around step 34, increasing the mean RMSE to 0.163. This discrepancy is an expected consequence of precision loss and non-associative floating-point arithmetic~\cite{golden2024flash, yuan2025understanding}.

Crucially, this inherent hardware-level divergence does not negatively impact end-to-end model fidelity. To confirm this empirically, we evaluate inference quality across understanding and generation tasks, as summarized in Table~\ref{tab:quality}. We report accuracy on MMAU-mini~\cite{sakshi2024mmau} and word error rate (WER) on LibriSpeech test-clean~\cite{panayotov2015librispeech} for understanding. For generation, we report WER on LibriSpeech test-clean and UTMOS on the Eval2000 dataset. While minor instance-level variations in synthesized speech naturally arise from hardware precision shifts and the inherent variance of TTS generation, outputs produced by our pipeline safely maintain the expected aggregate performance of the underlying models across all metrics.

\begin{table}[t]
    \centering
    \caption{Inference quality evaluation. Understanding is measured by MMAU-mini accuracy (\%, $\uparrow$) and ASR word error rate (\%, $\downarrow$). Generation is measured by TTS word error rate (\%, $\downarrow$) and dialogue UTMOS ($\uparrow$). ``--'' indicates the task is not supported by the model.}
    \label{tab:quality}
    \setlength{\tabcolsep}{4pt}
    \footnotesize
    \begin{tabular}{lcccc}
        \toprule
        & \multicolumn{2}{c}{\textbf{Understanding}} & \multicolumn{2}{c}{\textbf{Generation}} \\
        \cmidrule(lr){2-3} \cmidrule(lr){4-5}
        \textbf{System} & \makecell{\textbf{MMAU}\\[-2pt]\textbf{-mini}}$\uparrow$ & \textbf{ASR}$\downarrow$ & \textbf{TTS}$\downarrow$ & \textbf{UTMOS}$\uparrow$ \\
        \midrule
        Bagpiper (PyTorch)      & 74.5  & 2.5 & 2.7 & -- \\
        Bagpiper (vLLM)         & 75.4  & 2.6 & 3.3 & -- \\
        \midrule
        OpusLM (PyTorch)        & --  & 2.3 & 4.6 & -- \\
        OpusLM (vLLM)           & --  & 1.9 & 5.3 & -- \\
        \midrule
        OpusLM-Dial. (PyTorch)  & --  & -- & -- & 2.85 \\
        OpusLM-Dial. (vLLM)     & --  & -- & -- & 2.51 \\
        \bottomrule
    \end{tabular}
\end{table}



\subsection{CFG Overhead Analysis}
\label{sec:exp:cfg}

While CFG inherently doubles compute via parallel forward passes, Table~\ref{tab:cfg} shows our co-scheduling sustains over 80\% of baseline throughput. These measurements are conducted using Bagpiper on the LibriSpeech TTS task, operating at concurrencies of 256 and 512 for the CFG and baseline configurations, respectively. This efficiency stems from three factors. First, the companion request strictly copies the main sequence, ensuring near-perfect prefix cache hits and minimizing KV cache overhead. Second, since autoregressive decoding is fundamentally memory-bound, processing the companion request merely constitutes a marginal addition to the active batch size. This saturates idle compute to boost MFU with negligible latency and synchronization penalties. Finally, models like Bagpiper generate extensive textual reasoning before brief audio outputs. Selectively activating CFG solely during this short audio phase effectively amortizes the dual-pass cost across the entire end-to-end sequence.

\begin{table}[t]
    \centering
    \caption{CFG overhead on Bagpiper (vLLM). Prompt and generation throughput (tok/s) and MFU (\%) with and without Classifier-Free Guidance.}
    \label{tab:cfg}
    \small
    \begin{tabular}{lccc}
        \toprule
        \textbf{Setting} & \textbf{Prefill (tok/s)} & \textbf{Decode (tok/s)} & \textbf{MFU (\%)} \\
        \midrule
        w/o CFG & 315.8 & 4952.0 & 8.0 \\
        w/ CFG  & 271.2 & 3960.7 & 12.8 \\
        \bottomrule
    \end{tabular}
    \vspace{-10pt}
\end{table}

\section{Conclusion}
\label{sec:conclusion}

In this work, we presented a unified inference architecture that extends a continuous batching engine to natively support Speech Language Models. By integrating synchronous multi-stream sampling and delay-pattern de-interleaving into the generation loop, our design effectively handles the complexities of multi-layered audio tokens. Through a \emph{Paired Request Co-Scheduling} strategy, our system efficiently absorbs dual-branch overheads, sustaining 80\% of non-CFG throughput. Experimental results across multiple SpeechLM architectures demonstrate up to $108\times$ generation throughput improvement over sequential PyTorch inference while maintaining strict numerical correctness. As the modifications remain confined to the model and logit processing layers, this architecture provides a portable and high-performance foundation for real-time multimodal voice applications.

\section{Generative AI Use Disclosure}
Artificial intelligence tools were utilized exclusively for language editing and manuscript polishing. The core technical contributions, including the system architecture design, framework implementation, and experimental evaluation, were independently conducted by the authors.

\section{Acknowledgment}

This work used the Bridges2 at PSC and Delta/DeltaAI NCSA systems through CIS210014 from the ACCESS program, supported by NSF \#2138259, \#2138286, \#2138307, \#2137603, and \#2138296.

\bibliographystyle{IEEEtran}
\bibliography{mybib}

\end{document}